\documentclass[preprint,showpacs,showkeys,floatfix]{revtex4}

\usepackage{bm}
\usepackage{graphics}
\usepackage{graphicx}
\usepackage[dvips,usenames]{color}

\newcommand{\vs}{\vspace{4mm}}

\newcommand{\be}{\begin{equation}}
\newcommand{\ee}{\end{equation}}
\newcommand{\ba}{\begin{eqnarray}}
\newcommand{\ea}{\end{eqnarray}}

\newcommand{\AuthorTeam}{
 \author{T.~S.~Bir\'o %\footnote{On leave from KFKI RMKI Budapest, Hungary}
 } 
 \affiliation{
   \centerline{\hbox{KFKI Research Institute for Particle and Nuclear Physics,}}
   H 1525 Budapest, P.O.Box 49, Hungary
  }
  \author{G.~Kaniadakis 	
  } 
  \affiliation{
    Dipartimento di Fisica and Instituto Nazionale di Fisica della Materia,
    Politecnico di Torino, Corso Duca degli Abruzzi 34, 10129 Torino, Italy
  }
}
\begin{document}

\title{
Two generalizations of the Boltzmann equation
% On the equivalence of Kaniadakis and Biro extension of the Boltzmann eq.
%Equivalent non-extensive extensions of the Boltzmann equation
% Transformation between non-linearity and non-additive energy in kinetic models
}

\AuthorTeam

%\date{{\bf \today}}
% 25.75.Nq quark gluon plasma in heavy ion collisions
% 05.20.Dd classical kinetic theory
% 05.90.+m other topics in stat.phys. and thermodynamics 
% 02.70.Ns molecular dynamics and particle methods in computation


\begin{abstract}
 We connect two different extensions of Boltzmann's kinetic theory by
 requiring the same stationary solution. Non-extensive statistics
 can be produced by either using corresponding collision rates  
 nonlinear in the one-particle densities or equivalently by using
 nontrivial energy composition rules in the energy conservation 
 constraint part. Direct transformation formulas between key functions
 of the two approaches are given. 
% Models linear in density and complex
% in the energy expression require only an ${\cal O}(N^2)$ computing
% effort.  
\end{abstract}

\maketitle

%%%%%%%%%%%%%%%%%%%%%%%%%%%%%%

%%%%%%%%%%%%%%%%%%%%%%%%%%%%%%%%%%%%%%%%%%%%%%%%%%%%%%%%

\vs
Power-law distributions in nature are nearly as common as Gaussian
distributions, which is the limiting distribution emerging for the
properly scaled sum of infinitely many independent random variables. 
The power-law tailed distributions have escaped somehow the strength
of the central limit theorem: either by being composed from long-tailed
individual distributions, or by featuring finite-size effects in a
general sense. To the latter category belong unscreened long-range
forces and long-time memory effects; to the former a (multi)fractal
phase space occupation.
Since power-law tailed distributions can be found in many areas
from particle physics and astrophysics to financial market 
models\cite{MT-SCALING,T-TEAM-FITS,ASTRO},
it is righteously suspected that these belong to some 
more or less universal stationary
states of complex dynamicses. The question arises whether there are
characteristic common features in these dynamicses and what they 
then are.

\vs
Non-extensive thermodynamics has been developed in the past two decades
as a statistical theory to deal with such physical stationary 
states\cite{TSALLIS-ENTROPY,TSALLIS-RULES,TSALLIS-WANG}.
Initially based on mathematical investigations of a generalized definition
of Boltzmann's entropy, a never-decreasing macroscopical state-parameter
intimately connected to statistical probabilities of microstates, this
theory soon started to study dynamicses possibly leading to such
states. Studies of anomalous diffusion, random walk, and noisy 
equations\cite{MULT-NOISE,KODAMA,WILK}
revealed that either a nonlinearity in the one-particle probability or a
nontrivial interaction with the environment, which depends on the observed
low degree of freedom subsystem itself, may be responsible for such a
nontraditional behavior.

\vs
While the above approaches all mound in the study of an extended
Fokker-Planck problem, the set of all dynamical evolutions leading to
a thermodynamical state is much wider. In general a considerable
fluctuation of the intensive parameters may establish a non-Boltzmannian
canonical distribution\cite{SUPER-STAT}.  Another classical field of
establishing thermodynamics (and hence also non-extensive thermodynamics)
is kinetic theory in general. The basic assumption of Boltzmann, that
the time evolution of many systems has a fast, micro-reversible and
ergodizing component, which he comprised into a collision integral,
opens up studies of more complex non-linear dynamical theories.

\vs
Boltzmann's original kinetic theory is based upon a collision rate,
which is multilinear in the one-particle densities
and symmetric both in the colliding partners and against
time reversal. Based on these properties the H-theorem was derived,
funding the quantity entropy, and giving a way towards a
microscopic establishment of thermodynamics.
The Boltzmann entropy is an extensive (additive) quantity reflecting
the independence of microstate probabilities in weakly interacting
subsystems.
The classical Boltzmann equation 
restricts  possible stationary distributions to satisfy a
product rule, once an addition rule for energy is given.

\vs
Recent generalizations of the Boltzmann equation hook in exactly at this point:
either the product formula is generalized\cite{NLBE,HQ-THEOREM}, 
interpretable as mimicking
nontrivial many-body correlations in the equilibrium state, or the
addition of energy is replaced by a more general 
formula\cite{NEBE}, accounting for
an in-medium interaction energy shared by the colliding partners.
These two approaches may lead to the same, non-standard stationary
one-particle distribution. The aim of this work is to clarify
the interrelations between them and to present corresponding mathematical
formulas.

\vs
In Ref.\cite{NLBE} possible extensions
with nonlinear dependence of the collision rate on the one-particle
densities have been considered, 
this approach we shall refer here as the nonlinear Boltzmann
equation (NLBE). The rate in this approach consists of production
and blocking factors, both in gain and loss terms. Considering a simple
$ 1 + 2 \leftrightarrow 3 + 4$ two to two body collision, the rate of
change of the one-particle phase space density, $f_1=f(\vec{p}_1)$, is
given by
\be
 \dot{f}_1 \: = \: \int_{234}\!\!\!\! w_{1234} \left(a_3b_1a_4b_2 - a_1b_3a_2b_4 \right) 
\label{NLBE}
\ee
with $a_i=a(f_i)$ being the general production and $b_i=b(f_i)$ the blocking
factors. The transition probability rate factor, $w_{1234}$ may contain
a $1-3$ and a $2-4$ symmetric contribution also from correlations between
production and blocking. This is a general extension of the Boltzmann and the Boltzmann
Uhling-Uehlenbeck equations with respect to nonlinearity of the collision rate.
The standard theories are recovered for $a(f)=f$ and $b(f)=1$, or 
$b(f)=1\pm f$ respectively. In this case the stationary distribution is
governed by the ratio $\kappa(f)=a(f)/b(f)$ which becomes the traditional
Boltzmann factor:
\be
 \kappa(f_{eq}) = \exp( -E/T ).
\label{KAPPA}
\ee
This result assumes that in two-body collisions momenta and
energy are composed additively 
($E_1+E_2=E_3+E_4$,  $\vec{p}_1+\vec{p}_2=\vec{p}_3+\vec{p}_4$).
An H-theorem can be proven with the generalized expression for the entropy,
\be
 S_K \: = \: \int_1 \sigma(f_1) 
\label{K-ENTROPY}
\ee
where the integration is over the one-particle phase space. It turns out that 
$\sigma(f)$ is related to the previous quantities via
\be
 \sigma'(f) = - \ln \kappa (f).
\ee
Recently another approach to generalize Boltzmann's original treatment
has been proposed\cite{NEBE}. Here the (multi)linearity of the collision rate
is kept, blocking factors are not applied,
but the additivity of the energy during the micro-collisions
is replaced by a more general requirement: only a given function of
the individual energies, physically standing for the total two-particle
energy, is conserved, $h(E_1,E_2) = h(E_3,E_4)$.
The function $h(x,y)$ describes a general, non-extensive energy composition
rule for the two-body system. If this is chosen with the property of
associativity, $h(h(x,y),z)=h(x,h(y,z))$, then its most general form
is related to a strict monotonic function, $X(x)$:
\be
 h(x,y) \: = \: X^{-1} \left( X(x) + X(y) \right).
\label{COMPOSITION-RULE}
\ee
The function $X(x)$ is a mapping of the non-extensive composition rule
to the addition rule, it is unique up to a real multiplicative factor. 
The stationary distribution in this case is given by
\be
  f_{eq} \: = \: \exp\left( -X(E)/T \right).
\label{IKSZ}
\ee
The H-theorem can be proven for Boltzmann's original construction, but
this quantity may be interpreted as the total of the additive mappings
of non-extensive entropy contributions:
\be
 S_B \: = \: X_s(S_N) \: = \: - \int f \ln f.
 \label{B-ENTROPY}
\ee
By requiring a connection to the stationary state described by the
previous approach,
it turns out that the same composition rule should be
applied to the energy (scaled by the temperature) and to the entropy, 
i.e. $X(E)/T=X_s(E/T)$ with the respective mapping function for energy
and entropy.
The non-extensive Boltzmann equation
(NEBE) approach is still multilinear in the one-particle distributions.
In parton cascade simulations, by applying the above rules, 
the non-extensively composed total energy, 
$E_{tot}=X^{-1}\left(\sum_i X(E_i) \right)$, is conserved.

\vs
Of course, one may consider to apply both extensions, the nonlinear density
dependence and the non-extensive energy composition rule. In order to support
a given non-extensive thermodynamics, however, each one alone suffices.
In this respect these two generalizations are equivalent and transformation
formulas can be obtained between them. 
In the followings we review this correspondence.

\vs
A common stationary distribution relates the $\kappa(f)$ and the $X_s(E/T)$
functions:
\be
  f_{eq} = \kappa^{-1}(e^{-x}) \: = \: e^{-X_s(x)}
 \label{KAPPA-X-RELATION}
\ee
with the argument $x=E/T$. This allows for obtaining both $\kappa(f)$
or $X_s(E/T)$ to a given distribution, or converting these two basic
functions into each other. Based on this the non-extensive entropy
formulas (\ref{K-ENTROPY}) and (\ref{B-ENTROPY}) can be derived,
the relation
\be
 \sigma'(f) = - \ln \kappa(f) = X_s^{-1}(-\ln f)
 \label{LOG-KAPPA-X-RELATION}
\ee
is just a consequence thereof. Applying this relation to the Boltzmann entropy
(\ref{B-ENTROPY}) we arrive at
\be
 X_s(S_N) =  \int f X_s(s(f)),
\ee
with the one-particle contribution $s(f)=\sigma'(f)$
at the generalization of the mapping of the non-extensive entropy
to an additive entropy measure, $S_B$.
The basic entropy transformers of the two approaches, $\sigma(f)$ and
$X_s(E/T)$ must be related by (\ref{LOG-KAPPA-X-RELATION}) in order to
describe the same non-extensive thermodynamics in the stationary state.
As a trivial consequence the same composition rule, $h(x,y)$,  is applied to
the energy and to the entropy up to a scale factor $T$.

\vs
It is noteworthy that the Tsallis entropy can actually be obtained
by considering $S_T=\int f s(f)$. This very expression fulfills
an H-theorem in the case of Tsallis distribution, the corresponding
Tsallis and NLBE
$\sigma'(f)$-s being linear functions of each other (see below).
In the general case they are, however, different.
The deformed exponential and logarithm functions, considered in several
approaches to the non-extensive thermodynamics, are simply related to the
scaled mapping function\cite{DEFORM}:
\be
 \exp_{def}(t) \: = \: e^{-X_s(-t)}, \qquad \ln_{def}(f) \: = \: - X_s^{-1}(-\ln f).
\ee

%%%%%%%%%%%%%%%%%%%%%%%%%%%%%%%%  %%%%%%%%%%%%%%%%%%%%%%%%%%%%%%%

\vs
Finally we demonstrate the above relations for some well known
stationary distributions, relating them to specific non-extensive 
composition rules, production and blocking factors,  or expressions
for the total entropy, respectively.

\vs
The Maxwell-Boltzmann distribution, or in general the Boltzmann-Gibbs
distribution, $f(E)= \exp(-E/T)$ 
is recovered as a particular case in both approaches.
There is no blocking factor, $b(f)=1$, the production factor is linear,
$a(f)=f$. The mapping function is the identity, $X_s(t)=t$ and
the energy (and entropy) composition rule is the simple addition,
$h(x,y)=x+y$.

\vs
The Tsallis distribution, $f(E) = (1+aE)^{-1/aT}$  arises
for $X_s(x)=\frac{1}{a}\ln(1+ax)$\cite{ABE} by the energy composition rule
$h(x,y)=x+y+axy$\cite{TSALLIS-WANG}. The production to blocking factor ratio
becomes $\kappa(f)=a(f)/b(f)=\exp((1-f^{-aT})/aT)$.
The Tsallis index is given by $q=1-aT$. Note that if the parameter
$a$ in the generalized composition rule is temperature independent,
then the resulting Tsallis index will depend on temperature.
The NLBE entropy is given by $S_K= \frac{1}{1-q} \int (f^q/q-f)$,
the Tsallis entropy by $S_T=\frac{1}{1-q} \int (f^q-f)$.
This is not a principal difference, because whenever an H-theorem
is fulfilled with a given $\sigma'(f)$, a linear combination,
$A\sigma'(f)+B$ also fulfills the same H-theorem. The choice
$A=q$ and $B=1$ transforms $S_K$ into $S_T$ in this case.

Its mapping to an additive quantity coincides with the R\'enyi
entropy\cite{OTHER-ENTROPIES}, $S_R = X_s(S_T) = \frac{1}{1-q}\ln \int f^q$, 
when the distribution $f$ is normalized to one.

\vs
A variant of the L\'evy distribution, $f(E) =  \exp(-(E/T)^v)$,
with a fractional power $v$ can be achieved by using the
production to blocking ratio $\kappa(f) = \exp(-(-\ln f)^{1/v}) $
in the NLBE.
The mapping to additive quantity is done by $X(x)=x^v$, generating
the abstract composition rule: $h(x,y)=\left(x^v+y^v\right)^{1/v}$.

\vs
A pure power-law distribution, $f(E) =  (bE/T)^{\frac{1}{q-1}}$,
may be produced by using  $\kappa(f) = \exp(-f^{q-1}/b)$ or equivalently
by mapping the energy to an additive quantity via $X(x)=\frac{1}{1-q}\ln bx$.
The corresponding composition rule is given by $h(x,y)=bxy$.
The NLBE entropy becomes $S_K = \int f^q/bq$, the Tsallis entropy
is given by $S_T = \int f^q/b$ in this case. For $b=1$ they coincide.

\vs
The so called  quon distribution, $f(E) =  \frac{1}{e^{E/T}+q}$
(including the Fermi ($q=1$) and Bose ($q=-1$) distributions as particular cases),
is achieved by the production to blocking ratio: $\kappa(f) = \frac{f}{1-qf}$
(the choice $a(f)=f$ and $b(f)=1-qf$ leads to the Uhling-Uehlenbeck
equation).  The corresponding mapping to additive quantity is given by
$X(x)=\ln(q+e^x)-\ln(q+1)$
($X(0)=0$ is achieved by a subtraction of a finite constant.
 This does not work for $q=-1$ signaling the divergence due to
 Bose condensation.).
The generalized addition rule is somewhat complicated: 
$h(x,y)=x+y+ln\left(1+q(e^{-x}+e^{-y})+q(q-1)e^{-(x+y)} \right) -\ln(q+1)$
features a ''Pauli potential'', a pair energy leading to an
exact Fermi distribution. In the NEBE approach this can be achieved
without considering a third (blocking) particle at each collision. 
The density for the $S_K$ entropy is given by $\sigma(f)=-f\ln f - (1/q-f)\ln(1-qf)$.

\vs
Another distribution, proposed in Ref.\cite{KANI} for applying to a kinetic
theory of relativistic particles, applies the mapping function
$X(x)=\frac{1}{k} \cdot {\rm asinh}(kx)$. Another long-known formula, the mapping
of relativistic velocities to an additive rapidity variable, follows
from Einstein velocity-composition formula: $X(v)=c \cdot {\rm atanh}(v/c)$.

\vs
Some different distributions may be united in two- (or more) parameter classes.
For example, $X(x)=\frac{1}{aq}\left((1+ax)^q-1\right)$ for small $q$
but arbitrary large $x$ approaches Abe's logarithmic formula leading to
the Tsallis distribution, while for
large $a$ comes close to $X(x)=x^v$ leading to the L\'evy 
distribution\footnote{The corresponding deformed exponentials and logarithms are
then also two-parameter families.}.

\vs
Finally non-associative composition rules can be simulated in computerized
parton cascade simulations, too. In this case, however, no mapping can
be found to an additive, statistical quasi-energy. The individual energies
after a micro-collision, although random in a certain kinematical range, are no
more one-variable functions of the respective energies of the incoming pair,
but depend on both energies.
Non-associative composition rules, on the other hand, would not be
able to converge in the thermodynamical limit of repeated compositions of
compositions. Therefore they can probably be only of pure mathematical
interest.

%%%%%%%%%%%%%%%%%%%%%%%% RESULTS OF NUMERICS %%%%%%%%%%%%%%%%%

%%%%%%%%%%%%%%%% FIG.1  MOMS %%%%%%%%%%%%%%%%%

\begin{figure}

\begin{center}
\includegraphics[height=90mm,angle=-90]{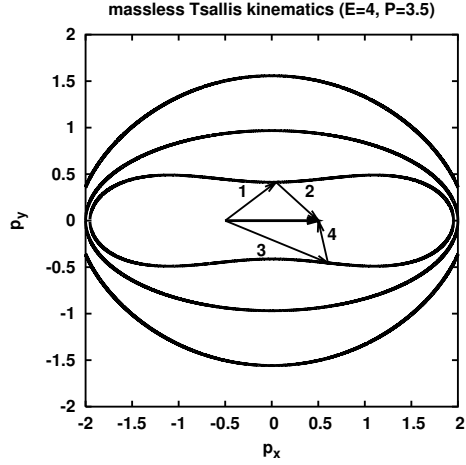}
\end{center}

\caption{\label{MOMS}
Kinetic constraint for momentum changes in a pairwise
elastic micro-collision using Tsallis rules for the energy composition
and massless free dispersion relation. The inner curve corresponds to
$a=0.125$, the middle one to $a=0$ and the outer one to $a=-0.125$.
}
\end{figure}

%%%%%%%%%%%%%%%%%%%%%%%%%%%%%%%%%%%%%%%%%%%%%

%%%%%%%%%%%%%%% FIG.2 DISTR %%%%%%%%%%%%%%%%%%%%

\begin{figure}

\begin{center}
\includegraphics[height=120mm,angle=-90]{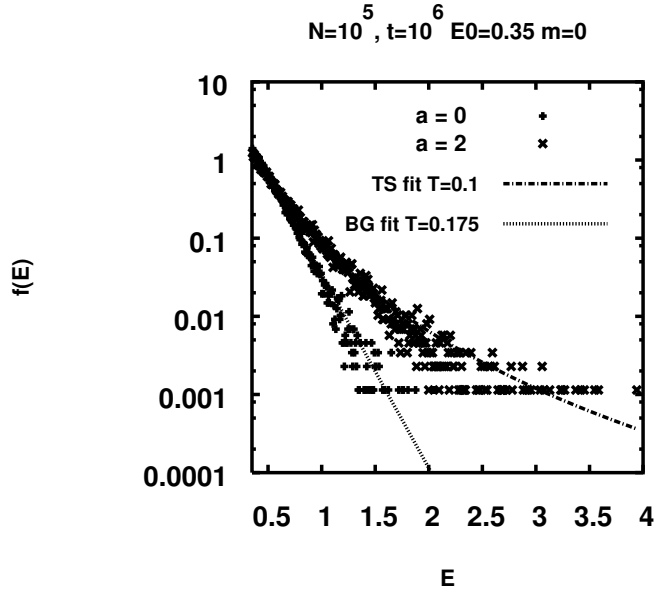}
\end{center}

\caption{\label{DISTR}
 Tsallis distributions obtained numerically by solving the corresponding 
 nonextensive Boltzmann equation.
}
\end{figure}

%%%%%%%%%%%%%%%%%%%%%%%%%%%%%%%%%%%%%%%%%%%%%%%%%

\vs
We have carried out numerical simulations so far for the Tsallis case\cite{NEBE}.
The nonextensive Boltzmann equation is simulated by $N=10^5-10^6$ test
particles having initially random momenta $\vec{p}_i$ in a certain range
and evolving by pairwise changes according to the rules
\be
 \vec{p}_1 + \vec{p}_2 = \vec{p}_3 + \vec{p}_4, \qquad
 X(E(p_1)) + X(E(p_2)) = X(E(p_3))+ X(E(p_4))
\ee
with the energy mapping $X(E)=\frac{1}{a} ln(1+aE)$. 
This simulation conserves the quasi energy $X(E_{tot})=\sum_i X(E(p_i))$.
Fig.\ref{MOMS} shows  pairs of momentum vectors satisfying the above
conditions for massless (extremely relativistic) particles, i.e. for
$E(p_i)=|\vec{p}_i|$. The two-dimensional section of the surface of
momenta  is an ellipse for $a=0$ (the traditional Boltzmann case) only,
for positive or negative values of this parameter a fourth order
curve is drawn. Final, stationary distributions of the one-particle
bare energy, $f(E)$ with $E=|\vec{p}|$ are then Tsallis distributions.
Fig.\ref{DISTR} shows examples of these distributions.
These are numerical demonstrations of how to achieve a Tsallis distribution
from arbitrary initial distributions.

%%%%%%%%%%%%%%%%%%%%%%%%%%%%%%%%%%%%%%%%% CONCLUSION %%%%%%%%%%%%%%%%%%%%%%%%%%%%
%\newpage

\vs
In conclusion we have demonstrated that
quite different ans\"atze to generalize Boltzmann's classical
kinetic theory produce the same stationary distributions of
non-extensive thermodynamics.
Upon this equivalence there always exist a strict monotonous mapping to an
additive entropy from more exotic entropy definitions, as long as
the energy composition rule applied in the micro-events (collisions)
are associative. It is natural to assume that the relevant rules
in the thermodynamical limit have to become associative.
The H-theorem holds.  Several known non-Boltzmann distributions belong to a
physically intriguing non-extensive energy addition rule.
This gives hope to find realizations in nature by studying
the pair interaction mechanism and its thermodynamical limit.
Nonlinear production and blocking factors may formally be
replaced by non-extensive energy formulas if only the
stationary distribution is asked for. The second method,
not using phase space blocking factors, allows for
computer simulations with a resource effort not worse than
${\cal O}(N^2)$.

\vs

%%%%%%%%%%%%%%%%%%%%%%%%%%% ACKNOWLEDGMENT %%%%%%%%%%%%%%%%%%%%%%%%%%

\vs
{\bf Acknowledgment}

Enlightening discussions with Drs. G\'eza Gy\"orgyi at E\"otv\"os University
and Antal Jakov\'ac at the Technical University Budapest are hereby
gratefully acknowledged.
This work has been supported by the Hungarian National Science Fund OTKA
(T049466) and the Deutsche Forschungsgemeinschaft.

%%%%%%%%%%%%%%%%%%%%%%%%%%% BIBLIOGRAPHY %%%%%%%%%%%%%%%%%%%%%%%%%%%%%

%%%%%%%% tex/notes/TSALLIS/QM-TSALLIS/bib.dat %%%%%%%%%%%
%
%	literature by topics to paper
%	T.S.Biro  07.01.2005.
%	updated for NEXT contribution, 15.09.2005
%
% ------------------------------------------------------

\end{document}